\documentclass[11pt]{article}
\usepackage[letterpaper, left=1in, top=1in, right=1in, bottom=1in, verbose, ignoremp]{geometry}
\pdfoutput=1
\RequirePackage[OT1]{fontenc}
\RequirePackage{amsthm,amsmath,amsfonts,amssymb}
\RequirePackage{natbib}
\usepackage[hyphens]{url}
\RequirePackage[colorlinks,citecolor=blue,urlcolor=blue]{hyperref}

\usepackage{bm,graphicx,subfigure,array,enumerate,caption}
\usepackage{multirow,multicol}

\usepackage{tikz}
\usetikzlibrary{patterns}

\def\beginmat{ \left( \begin{array} }
\def\endmat{ \end{array} \right) }

\def\log{{\rm log}}

\newcolumntype{P}[1]{>{\centering\arraybackslash}p{#1}}
\newcolumntype{L}{>{\centering\arraybackslash}m{8cm}}

\date{}

\newcommand{\bE}{\mathbf{E}}

\newtheorem*{theorem*}{Theorem}

\newtheorem*{claim*}{Claim}
\theoremstyle{definition}

\theoremstyle{remark}

\numberwithin{equation}{section}
\theoremstyle{plain}

\begin{document}

\begin{flushleft}
{\Large{\textbf{Classical Music Composition Using State Space Models}}}
\newline
\\
Anna K. Yanchenko\textsuperscript{1},
Sayan Mukherjee\textsuperscript{1,2}

\medskip

1 Department of Statistical Science, Duke University, Durham, NC 27708
\\
2 Departments of Computer Science, Mathematics, and Biostatistics \& Bioinformatics, Duke University, Durham, NC, 27708 \\
\medskip
*Corresponding Email: \url{aky5@stat.duke.edu}
\end{flushleft}

\date{}

\bigskip

\begin{abstract}
Algorithmic composition of music has a long history and with the development of powerful deep learning methods, there has recently been increased interest in exploring algorithms and models to create art.  We explore the utility of state space models, in particular hidden Markov models (HMMs) and variants, in composing classical piano pieces from the Romantic era and consider the models' ability  to generate new pieces that sound like they were composed by a human.  We find that the models we explored  are fairly successful at generating new pieces that have largely consonant harmonies, especially when trained on original pieces with simple harmonic structure.  However, we conclude that the major limitation in using these models to generate music that sounds like it was composed by a human is the lack of melodic progression in the composed pieces. We also examine the performance of the models in the context of music theory.
\end{abstract}
\vspace*{.2in}

\section{Introduction}
\label{sec:intro}

The objective of constructing algorithms to compose music or composing music with minimal human intervention has a long history. Indeed in the 18th century a musical game called
Musikalisches W\"urfelspiel was developed that took small music fragments and  combined them randomly by chance, often by tossing dice \citep{cope:96a}. The first musical composition
 completely generated by a computer was the  ``Illiac Suite", produced from 1955 - 1956 \citep{Nierhaus:2009}. A wide variety of probabilistic models and learning algorithms have been 
 utilized for automated composition, including hidden Markov models (HMMs) to compose jazz improvisations, generative grammars to generate folk song rounds and artificial neural networks to harmonize existing melodies \citep{Nierhaus:2009}. Additional methods of algorithmic composition include the use of transition networks by \cite{Cope:1991} and multiple viewpoint systems by \cite{Whorley:Conklin:2016}.  \cite{Dirst:Weigend:1993} explored data driven approaches to analyzing and completing Bach's last, unfinished fugue and   \cite{Meehan:1980} discussed efforts to develop models for tonal music theory using artificial intelligence. 
 
 Artificial neural networks and deep learning methods have recently been used for
 automated music generation. Specifically, recurrent neural networks (RNNs) were used for algorithmic composition by  \cite{Mozer:1994} and \cite{Eck:Schmidhuber:2002}  and 
\cite{Boulanger-Lewandowski:EtAl:2012} used variants of RNNs for modeling polyphonic (multiple-voice) music. \cite{Hadjeres:Pachet:2016} developed the interactive DeepBach system, that used Long-Short Term Memory (LSTM) units to harmonize Bach chorales with no musical knowledge built into the system.  \cite{Johnson:2015} explored the use of RNNs for composing classical piano pieces.  There was recently a concert in London featuring work composed by RNNs that sought to  emulate the styles of known composers  \citep{Vincent:2016}.
Lastly, Google's Magenta project \citep{Magenta,MagentaDevSummit:2017} has been focused on generating art and music through deep learning and machine learning and maintains an open source code repository and blog detailing their efforts. 

In this paper we explore using probabilistic time series models, specifically variations of hidden Markov models and time varying autoregression models, for algorithmic composition of piano pieces from the Romantic era. The decision to focus on the specific probabilistic time series models over RNNs is the ease of implementing the probabilistic time series models as well as
their ability to model complex time series. The decision to focus on piano pieces from the Romantic era allows us to develop automated metrics on how well the compositions generated compare to the original pieces with respect to originality, musicality, and temporal structure. It would be very challenging to develop metrics that would make sense across the wide variety of musical genres. 

While HMMs have also been utilized for a variety of musical analysis and algorithmic composition tasks, most applications have focused on music classification and the harmonization of melodic lines, with applications to the composition of melodies more limited.   A survey of previous work using Markov models for music composition appears in \cite{Ames:1989}.  More recently, \cite{Suzuki:Kitahara:2014} used Bayesian networks to harmonize soprano melodies, \cite{Weiland:EtAl:2005} modeled the pitch structure of Bach chorales with Hierarchical HMMs, \cite{Pikrakis:EtAl:2006} used variable duration HMMs to classify musical patterns, \cite{Allan:Williams:2004} used HMMs to harmonize Bach chorales and \cite{Ren:2010} explored the use of HMMs with time varying parameters to model and analyze musical pieces.

The main goal and contribution of this paper is to use classical probabilistic time series models  to generate piano pieces from the Romantic era that sound like they were composed by a human, and to develop metrics to compare the generated compositions with the originals with respect to originality, musicality, and temporal structure. Specifically, we explore the dissonance
or consonance of the harmonies produced by the generated piece as compared to the original piece and the extent of global structure and melodic progression in the generated pieces.

The paper is organized as follows. In Section~\ref{sec:HMM}, we discuss the representation of the music used in our models, specify the probabilistic models, and describe the musical pieces used to train the models. In Section~\ref{sec:music_theory}, we describe the metrics we developed to assess the quality of the automated compositions. Finally, in Section~\ref{sec:results}, we analyze each model in terms of music theory and discuss the quality of the automated compositions using both the quantitative metrics we developed and a qualitative survey of listeners, including musicians. We also relate the performance of the generated compositions to some ideas in music theory. We close with conclusions and future work.

\section{Methods and Models}
\label{sec:HMM}
 
Automated generation of compositions requires a representation of the music that can be generated by time series models, specifically variations on 
hidden Markov models (HMMs). In this section we specify the representation of the music we used, the 14 time series models we used to generate compositions,
and the 10 piano pieces from the Romantic period we used as training data.

\subsection{Music Representation}
\label{subsec:musicrep}

\begin{figure}[ht]
\begin{center}
\begin{tabular}{c}
\includegraphics[height=20mm]{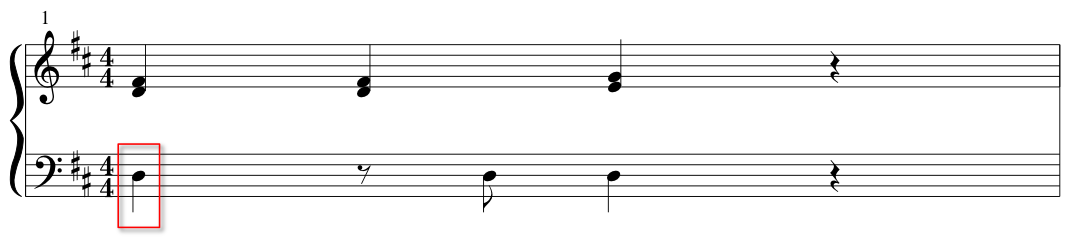} \\ \\
\begin{tabular}{|c| c|c |c|c|c|c|c|c|c|c|} 
 \hline
50&62&66&50&50&62&66&64&67&50&50 \\
\hline
\end{tabular}
\end{tabular}
\end{center}
\caption{The first bar of sheet music from Beethoven's Ode to Joy is displayed at the top. The equivalent symbolic representation for the first bar of Beethoven's Ode to Joy is displayed below. A specific note's duration is the difference in time stamps from the first occurrence of that pitch (the note is ``turned on") to when the same note pitch next occurs (the note is ``turned off").  For example, the first note pitch is 50 (corresponding to a D) and next occurs three pitches and one time stamp later, indicating that this D has a duration of one quarter note (highlighted in red above).} 
\label{notesheetex}
\end{figure}

Two standard representations of compositions are sheet music and the Musical Instrument Digital Interface (MIDI) format. MIDI is a data communications protocol that allows for the exchange of information between software and musical equipment through a digital representation of the data that is needed to generate the musical sounds \citep{Rothstein:1992}.  Sheet music is the standard representation used by composers and musicians, by which musical pieces are actually performed. The automated compositions we generate can be output in both MIDI as well as sheet music formats. However, we do not use the MIDI or sheet music formats in our statistical models, either in generating compositions or for learning compositions from example pieces. We consider a composition as a symbolic sequence of note pitches  over discrete fixed time steps $\{X_1,...,X_T\}$ where the discrete time steps are usually one sixteenth note or one eighth note in length and the pitches are encoded as an integer from $0-127$.  This representation allows us to model both single notes as well as chords.  See \autoref{notesheetex} for an example
of transforming sheet music to our pitch  representation. 
There is open source code \citep{midicsv} to convert from MIDI formats to our pitch representation and back.  \cite{midicsv} provides additional explanation about the exact format of the MIDI and CSV converted files.  GarageBand \citep{GarageBand} software is used to convert pieces in MIDI format to the equivalent representation in sheet music.

\subsection{Time Series Models}
\label{subsec:over}

The time series models considered in this paper are all variations of HMMs \citep{Rabiner:1989} with the observable variables  $\{X_1,...,X_T\}$ 
given by our pitch representation. We will denote an observed sequence $\{x_1,...,x_T\}$ as $x_{1:T}$. The likelihood for the basic HMM is
\begin{equation}
p(x_{1:T} \mid z_{1:T}, \theta,\theta',\pi) = \pi(z_1) \, p_{\theta}(x_1 \mid z_1) \, \prod_{t=2}^T p_{\theta'}(z_t \mid z_{t-1}) \, p_\theta(x_t \mid z_t),
\end{equation} 
where $z_{1:T}$ are the hidden states and the model parameters $\theta$ specify the emission probabilities $p_\theta(x \mid z)$, $\theta'$ specify the transition probabilities $p_{\theta'}(z_t \mid z_{t-1})$, and the initial distribution is $\pi$. The standard approach to infer model parameters is the Baum-Welch Algorithm \citep{Baum:1972}.

In addition to the standard HMM, we considered a variety of statistically richer Markov models. More complicated and longer range dependence structures were modeled using
higher-order Markov models (k-HMMs, \citep{Mari:Schott:2001}), where k specifies the order of dependence, and  Autoregressive HMMs (ARHMMs, \citep{Rabiner:1989,Guan:EtAl:2016}). Time varying transition probabilities were modeled 
via Left-Right HMMs  (LRHMMs) and hidden semi-Markov models (HSMMs, \citep{Yu:2010,Murphy:2002}). More complex hierarchical state structure was modeled using
two  state hidden Markov models (TSHMMs), factorial HMMs (FHMMs, \citep{Ghahramani:Jordan:1997}), and Layered HMMs (LHMMs, \citep{Oliver:EtAl:2004}).  Time-Varying Autoregression (TVAR) models were explored to model the  non-stationarity and  periodicity of the note pitches for the original pieces. 

{\em TVAR} models \citep{Prado:West:2010} are able to model non-stationarity and longer dependence between note pitches over time.  The TVAR model for a sequence $(X_1,...,X_T)$ is specified as
$$X_{t} = \sum_{j=1}^d \theta_j ((t-1)/T) X_{k-j} + \sigma(t/T)  \varepsilon_t, \quad t=1,...,T$$
where the variance $\sigma(t/n)$ is time varying,  $\varepsilon_t$ can also be time varying, and  the parameters $\theta_j ((t-1)/T)$ are also time varying.
We used forward filtering to fit each TVAR model and backwards sampling \citep{Prado:West:2010}  to generate new pieces. The order and discount factors of the TVAR models
were selected by a grid-search to maximize the  marginal likelihood. The orders of the TVAR models ranged from 7 to 14, depending on the training piece. Since the note pitch values were discrete (rather than continuous) and only certain note pitches occurred in the original piece, the generated note pitches were binned to the nearest pitch that occurred in the original piece.

We considered 15 models, {\em M1-M14}, to generate compositions: model {\em M1} was a standard HMM with 25 hidden states, model {\em M2} was a 2-HMM with 25 hidden states, 
{\em M3} was a 3-HMM with 10 hidden states,  model {\em M4} was a LRHMM with 25 hidden states, model {\em M5} was a 2-LRHMM with 25 hidden states, 
model {\em M6} was a 3-LRHMM with 10 hidden states, model {\em M7} was a ARHMM with 25 hidden states, model {\em M8} was a HSMM with 25 hidden states, 
{\em M9}  was a TSHMM with 10 hidden states in the first layer and 5 in the second layer, {\em M10} was a TSHMM with 5 hidden states in the first layer and 10 in the second layer, model 
{\em M11} was a FHMM with three independent HMMs each with 15, 10 and 5 hidden states respectively, model {\em M12} was a LHMM with three layers and 25 hidden states in each layer, model {\em M13} was a TVAR with order between 7 and 14  and model {\em M14} was a baseline model specified by a HMM with randomly assigned parameters. Parameter inference for all  the models except {\em M13}  and {\em M14} was executed by the appropriate Baum-Welch algorithm, run until convergence. 

\subsection{Romantic Era Compositions}
\label{sec:music}

All 14 models were trained on ten piano pieces from the Romantic period, downloaded from \cite{mfiles.co.uk,piano-midi.de} and \cite{midiworld}. The pieces were in MIDI format which was converted to our music representation of note pitches. In addition to the note pitch, there was a relative time-stamp associated with each observation.  The time stamps at which note pitch observations occurred were largely regular in time. However, to simplify modeling, we assumed all observations equally spaced in time. Additionally, some time stamps had multiple note pitches occurring at the same time, indicating a musical chord. However, we assumed that the note pitches were a univariate time series and treated each observation as sequential in time, whether or not subsequent observations occurred at the same time-stamp.   

All training pieces considered were either originally composed for piano or were arranged for piano and were selected to have a range of keys, forms, meters, tempos and composers.  The Romantic period of music lasted from the end of the eighteenth century to the beginning of the  twentieth  century.  This period was marked by a large amount of change in politics, economics and society.  In general, music from the Romantic era tended to value emotion, novelty, technical skill and the sharing of ideas between the arts and other disciplines \citep{Warrack:1983}.  The ten original training pieces, as well as their keys and time signatures are listed in \autoref{table:pieces} below.  

\begin{table}[h!]
\centering
\caption[Romantic era training pieces]{\raggedright Summary of the Romantic era piano training pieces modeled, including the composer, key signature,  time signature (TS) and empirical entropy (Ent.) of each piece. (M) refers to a Major key, while (m) corresponds to a minor key.}
\begin{tabular}{|c|p{8.5cm} |c|c|c|} 
 \hline
 Composer & Piece & Key & TS  & Ent. \\ [0.5ex] 
 \hline\hline
 Beethoven & Ode to Joy (Hymn Tune)  & D (M) & 4/4 &  2.328 \\\hline
Hopkins & We Three Kings (Hymn Tune) &   G (M) & 3/4 & 2.521\\\hline
Chopin & Piano Sonata No. 2, 3rd Movement (Marche funebre) & B$\flat$ (m) & 4/4  & 2.759\\\hline
Mendelssohn & Hark! The Herald Angels Sing (Hymn Tune) & F (M) &  4/4 & 2.780 \\\hline
 Mendelssohn & Song without Words Book 5, No. 3 & A(m) & 4/4 & 2.897 \\ \hline
 Beethoven & Piano Sonata No. 14 (Moonlight Sonata), 1st Movement & C\# (m)  & 3/4 & 3.000 \\ \hline
  Tchaikovsky & The Seasons, November - Troika Ride &  E (M) & 4/4 &  3.063 \\\hline
 Mendelssohn & Song without Words Book 1, No. 6 & G (m) & 6/8 & 3.227\\\hline
 Liszt & Hungarian Rhapsody, No. 2 & C\# (m) & 4/4 & 3.436 \\\hline
  Mussorgsky & Pictures at an Exhibition, Promenade - Gnomus & B$\flat$ (M) & 6/4 & 3.504 \\
 \hline
\end{tabular}
\label{table:pieces}
\end{table}

\section{Evaluation Metrics}
\label{sec:music_theory}
In this section we will propose a set of metrics to compare the similarity of the generated compositions with the original compositions in terms of originality, musicality, and temporal structure. There are two concepts in music theory that our metrics will need to capture: the concept of harmony and the concept of melody; see \cite{Laitz:2003} and \cite{Gauldin:2004} for details on musical theory. Melody is a sequence of tones  that are perceived as a single entity and is often considered a combination of pitch and rhythm. We will not summarize the music theory behind melody in this paper as our metric to capture melody or temporal structure will not require theory. On the other hand we will require some knowledge of the music theory underlying harmony to describe the metrics we use for musicality. Harmony refers to the arrangement of chords and pitches and the relation between chords and pitches.

\subsection{Harmony}

Romantic era music is  tonal, meaning that the music is oriented around and pulled towards the tonic pitch.  For this paper, we primarily consider a high-level view of the intervals and chords present in the generated pieces.  Briefly, there are two types of musical intervals, the melodic interval, where two notes are sounded sequentially in time, and the harmonic interval, where two or more notes are sounded simultaneously. Intervals can be either consonant or dissonant and  consonant intervals can further be categorized as perfect or imperfect consonances.  

Consonant intervals are stable intervals that do not require a resolution.  Perfect consonances are the most stable intervals, while imperfect consonances are only moderately stable.  However, neither type of consonance requires a resolution.  Dissonant intervals, on the other hand, are unstable and in Romantic era music need to be resolved to a stable, consonant interval.  Dissonant intervals sound incomplete and transient, and through the resolution to a stable, consonant interval, the phrase sounds complete.

In Romantic era music, very few, if any, dissonant intervals are left unresolved.  Dissonance serves to add interest and variation to the music and is often used to build tension in a phrase or melody.  The resolution to a consonant interval relieves this tension and restores the phrase to stability.

A chord is a combination of three or more different pitches, and the predominant chords occurring in the Romantic era were the triad and the seventh chord.  The triad is composed of a root pitch and the third and fifth above the root, while a seventh chord is the root, third, fifth and seventh.  Chords can be closed or open depending on the spacing between pitches in the higher and lower registers. 

For additional discussion of music theory, see Section 2 in the Supplementary Materials.

\subsection{Metrics}\label{sec:metrics}

The metrics we used in this paper capture how well the generated pieces conform to characteristics of the Romantic era. The metrics fall into three broad categories: originality, musicality and temporal structure.  The originality metrics measured how complex the generated pieces were, as well as how different the generated pieces were from the original training pieces.  The musicality metrics primarily attempted to measure the harmonic aspects of the generated pieces, while the temporal structure metrics attempted to measure the melodic qualities of the generated pieces. 

{\em Originality Metrics:}  The three originality metrics are based on information theory and distances between strings. The first metric is the empirical entropy of a musical piece, either
a training piece or a generated piece. Empirical entropy was proposed by \cite{Coffman:1992} as a measure of musical predictability.  Simple hymn tunes like  Beethoven's Ode to Joy have lower empirical entropy than complex, unpredictable pieces like Liszt's Hungarian Rhapsody, No. 2. The entropy of the training pieces are given in \autoref{table:pieces}. 

The second metric was mutual information \citep{Cover:2006:EIT:1146355}, which captured the relative entropy of the generated piece with respect to the training piece. The idea is that generated pieces that have
greater mutual information with respect to the training pieces are considered less original. For a given statistical model {\em M} we compute the mutual information between the posterior of the training sequence $p(x_1,...,x_T \mid M)$  and the generated sequence $p(y_1,...,y_T \mid M)$. The minimum edit distance is also a metric of dissimilarity between the generated sequence and the training sequence; in this case the metric is not based on a probability model. The minimum edit distance is the minimum number of insertions, deletions and substitutions 
necessary to transform one string into another \citep{Jurafsky:Martin:2009}.

{\em Musicality Metrics:} We considered three metrics to capture harmonic aspects of the generated pieces.   The first metric was a count of dissonant  melodic and harmonic intervals, normalized by the length of the piece.  The  minor second, major second, minor seventh and major seventh intervals are considered  dissonant \citep{Gauldin:2004}.  We expected the amount of dissonance in the generated pieces to be similar to the amount of dissonance in the original training piece.  


 We also considered the distribution of note pitches. The occurrence of each unique pitch throughout the length of the piece was counted and normalized by the number of total notes in the piece.  We expected  pitches that were used less in the original piece to also be less prevalent in the generated pieces.

{\em Temporal Structure Metrics:} We used measures of decay correlations in a time series to capture the amount of temporal structure in a musical piece, a metric that captured a sense of melody. The first metric we considered was the  autocorrelation function (ACF) of a sequence \citep{Prado:West:2010}. The ACF is the correlation coefficient between an observation at time $t$ and at a lag $t+h$
$$\gamma(h) =  \mbox{Corr}(x_{t+h}, x_t).$$
The idea behind the second metric considered, the partial autocorrelation function (PACF) \citep{Prado:West:2010}, is to measure the correlation between  $x_t$ and $x_{t-h}$ once the influence of $x_{t-1},...,x_{t-h+1}$ have been removed.  For example, in a Gaussian setting, the PACF can be calculated as:
$$\pi(h) =  \mbox{Corr}(x_t - \bE(x_t \mid x_{t-1},...,x_{t-h+1}),   x_{t-h} - \bE(x_{t-h} \mid x_{t-1},...,x_{t-h+1})).$$
A generated piece with a high degree of global structure and melody would be expected to have  ACF and PACF plots with some structure out to high lags.  The ACF and PACF was calculated for each generated piece out to lag 40. ACF and PACF plots for Beethoven's Ode to Joy are given in \autoref{ode}.

\paragraph{Evaluation}
After the considered model converged, 1000 new pieces were sampled from the learned model using the appropriate generative description of each model and the root mean squared error (RMSE) for each metric (except mutual information and edit distance) was calculated.   The RMSE was primarily used to rank the generated pieces, to select the top generated pieces for evaluation by human listeners and to gain insight into some general trends observed in the generated pieces.  The RMSE can be calculated as 
\begin{equation}
RMSE = \sqrt{\dfrac{1}{n}\sum_{i=1}^n \left(y_i - y_0\right)^2}
\end{equation}       
\noindent where $y_i$ is the considered metric, $y_0$ is the value of the metric for the original piece and $n$ is the number of generated pieces, in this case $n=1000$.

The RMSE was calculated for the musicality and temporal structure metrics, as well as for the empirical entropy, but not for the mutual information or minimum edit distance.  For the mutual information and minimum edit distance, both metrics were calculated for each of the 1000 generated pieces with respect to the original training piece, then the average of these 1000 values was taken to use for comparison between models and training pieces, in lieu of the RMSE.

\section{Results}\label{sec:results}
The quality of the generated compositions was evaluated in terms of the learned model parameters, the quantitative metrics developed and human listening evaluations.  First, a simple case study looking at ``Twinkle, Twinkle, Little Star" was considered as a simulation study to explore the learned model parameters and general traits of each model considered.  These trends were validated with a more complicated piece by Bach.  Then, the quantitative metrics calculated on the Romantic era training pieces were used to inform a human listener evaluation to explore subjective and qualitative aspects of the generated pieces.   

\subsection{Case Study, Model Inference and Validation}\label{sec:case_study}
In order to explore the learned model parameters in terms of music theory, we considered the simple tune ``Twinkle, Twinkle, Little Star" as an initial case study.  Twinkle, Twinkle, Little Star was selected for its simplicity in terms of melody, harmony and rhythm, allowing for an easier evaluation of model parameters.  After exploring the learned model parameters on this simple piece, we considered Bach's more complex Fugue No. 2 in C Minor, BMV 871 from the second book of The Well-Tempered Clavier to validate these observed trends.


\subsubsection{Twinkle, Twinkle, Little Star}
We considered a version of  ``Twinkle, Twinkle, Little Star" in the key of C major.  Twinkle, Twinkle, Little Star is a tonal piece, with the main, stationary harmonies consisting of the major triad and intervals of the perfect fourth and the perfect fifth.  The harmony primarily consisted of perfect consonances and there were no dissonant harmonic intervals in the original piece.

The simplest HMM with five hidden states was considered in depth and the initial and emission distributions, transition matrix and hidden states were explored and interpreted. Results of additional models are summarized in Section 3 in the Supplementary Materials.  In order to interpret the models and results in terms of music theory, we looked to the learned parameters, in particular the emission distributions, to determine trends involving the observed note pitches.  We then analyzed if these trends in the emission distributions corresponded to music theoretic concepts, that could then be mapped onto the hidden states, transition matrix and initial distribution. 

\paragraph{Representation of Harmony}
Our representation of the observed note pitches treated the original training piece as a univariate time series.  Since this representation lost information about the chords in the original piece, we expected the hidden states of the model to pick out the \textbf{harmonies} that occurred in the original piece. 

Harmony represents the underlying dynamics of the piece, and when a certain harmony occurs, only certain notes can express that harmony.  Thus, we looked to the emission distribution of the first order HMM to analyze if the hidden states were indeed representing the harmonies of the piece.  For example, if a specific hidden state represented the perfect fifth,  we would only expect the tonic note (C) and the perfect fifth (G) to be the observed notes, so the emission distribution would have a high probability for C and G for this specific hidden state. The transition matrix would then correspond to the transition between different harmonies.

\paragraph{Case Study Results}
The HMM with 5 hidden states was able to capture all of the predominant harmonies in the original piece.  Each of the main harmonies in the piece was represented in at least one hidden state (where the observed notes in the harmony had high emission probability for that state).  The five hidden states  represented the tonic pitch (C), the perfect fourth (F) and the major triad (C, E, G).  Overall, the emission distribution was fairly sparse.   

Interestingly, while the pitches A and D occurred with some frequency in the original piece, these secondary harmonies were never observed with the highest emission probability across all of the hidden states, indicating that these harmonies were not well modeled by the HMM.  The transition matrix showed a fairly equal distribution of probabilities amongst the hidden states, representing the dynamics of the harmonies in the piece.  The hidden state corresponding to the tonic pitch (C) had the highest initial distribution probability, which corresponded to the fact that the original piece began on the tonic (C).  These results held as the number of hidden states was further increased and were also observed in the other models considered.  

In summary, the hidden states of the HMM represented the predominant \textbf{harmonies} in the original piece, even though the observed states were  treated as a univariate series.

\subsubsection{The Well-Tempered Clavier}
To validate our findings above on a more complicated piece, we considered Bach's Fugue No. 2 in C Minor, BMV 871 from the second book of The Well-Tempered Clavier.  Pieces by Bach are frequently used for algorithmic composition tasks.  This piece was much more complicated than Twinkle, Twinkle, Little Star, in terms of harmony, melody and rhythm.  Focusing on the harmonies in the original piece, the piece was in C minor and contained several accidentals (note pitches not in the key signature), leading to more complex harmonies.  Looking at the emission distribution for an HMM with 5 hidden states, all accidental notes had very low emission probabilities, indicating that when the harmonies were modulated in the original piece, the HMM was not able to model these harmonic subtleties.  This also created a more dissonant generated piece.

As observed for Twinkle, Twinkle, Little Star, each hidden state emitted observed notes with highest emission probabilities that corresponded to predominant harmonies in the original piece.  For example, one hidden state emitted G and E$\flat$, which combine with C to form the minor triad.  Another hidden state had F, G and C as the observed notes with the highest emission probabilities, where F is the perfect fourth of C and G is the perfect fifth of C.  These harmonies occurred frequently in the original piece by Bach.

The transition matrix was sparse, where each hidden state transitioned to only one other hidden state with high probability.  Finally, the initial distribution had probability one for the hidden state which emitted G and E$\flat$ with the highest probabilities.  E$\flat$ is the relative major of C minor, so while not being the tonic, this pitch was important to the harmonies in the original piece.  

Overall, for the more complex piece of Bach's Fugue No.2 in C Minor, BMV 871, we observed similar trends in terms of the hidden states and parameters of the first order HMM as seen for the simpler Twinkle, Twinkle, Little Star.


\subsection{Results on Compositions}

Our overarching goal was to generate new compositions that recapitulated the several well-defined harmonic and melodic characteristics of Romantic music. We were also interested in two questions about automated compositions: (a) which time series models are the best at  generating Romantic era compositions and (b) which original compositions are more amenable to generating realistic Romantic era pieces and does music theory offer some insight into what properties of these pieces may drive variance. Ultimately, judging the quality of a generated composition  requires evaluation by human listeners. However, it is not feasible to provide pieces generated by all models and all original pieces to listeners, so we used the quantitative metrics described in the previous section to select the top compositions that were presented to human listeners.

The Romantics valued emotion and virtuosity in their music so the generated compositions needed to elicit some emotional response in human listeners.  To evaluate  subjective elements of the generated pieces, such as Romantic era style and the human-like qualities of the generated pieces, several human listeners of varying musical backgrounds listened to and evaluated several generated pieces. This was the ultimate test of our generated compositions.

We first outline the evaluation procedures for both the numerical metrics and the human listeners. We then discuss in detail analysis and summaries that arose from the numerical and human evaluations. We close the section by relating some of our observations from the evaluations to music theory.

\paragraph{Numerical Evaluation} Compositions were generated for all ten original pieces and models  {\em M1}-{\em M12} and {\em M14}. Given the large number of compositions (140), the purpose of the numerical evaluation was to select the top compositions with respect to originality, musicality or harmony, and temporal structure or melody. The metric we used to assess originality was the minimum RMSE empirical entropy. The metrics we used to  assess harmony and melody were the averages of the  RMSE of the musicality metrics and the temporal structure metrics, respectively.

\paragraph{Listening Evaluation}  

The top three pieces were evaluated by sixteen human listeners of varying musical backgrounds. Eight of the individuals were currently in a musical ensemble and were considered ``more" musically knowledgeable, while the other eight individuals were not currently in a musical ensemble. Each individual was told that  all three of the pieces had been generated by a computer using statistical models trained on an original piano piece from the Romantic era  The pieces were labeled A, B and C, so the listeners did not know a priori the original training piece for each generated piece.  In addition to ranking the three pieces in order of their favorite to least favorite, each individual was asked:  \textit{What did you like and not like about each piece? Did any of the pieces sound like they were composed by a human? Why or why not?}

\subsection{Analysis of Numerical Evaluation}

The top three pieces according to the numerical metrics were Chopin's  Piano Sonata No. 2, 3rd Movement (Marche funebre) modeled by a layered HMM (lowest temporal structure RMSE), Mendelssohn's Hark! The Herald Angels Sing modeled by a layered HMM (second lowest empirical entropy RMSE, Chopin piece already selected) and Beethoven's  Ode to Joy modeled by a first order HMM (lowest musicality metrics RMSE), see \autoref{table:eval_RMSE} for the RMSE scores. These three pieces were used for the human evaluations. 
\begin{table}[h!]
\caption[Entropy summaries]{ \raggedright Summary of the entropy RMSE, musicality average RMSE (Music), and temporal structure average RMSE (Time) values for listening evaluation pieces.}
\centering
\vspace{-.2em}
\begin{tabular}{ |c|c|c|c|c| }
 \hline
Training Piece & Model & Entropy  & Music  & Time \\\hline\hline
Ode to Joy & First Order HMM      &      0.042 &  0.018 & 0.142\\\hline													
Marche funebre &     Layered HMM & 0.021 &   0.019 & 0.049	\\\hline						
Hark! The Herald Angels Sing & Layered HMM &   0.022 &  0.027 &  0.075\\\hline						
\end{tabular}
\label{table:eval_RMSE}
\end{table}

\subsection{Analysis of Human Evaluation}

\begin{figure}[ht]
\centering
\subfigure[]{
\includegraphics[width=0.45\textwidth]{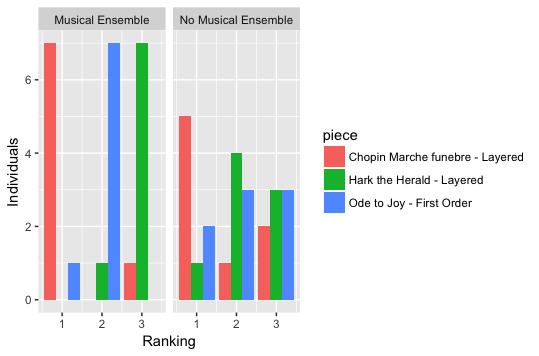}
\label{rank_music}
} \hspace{.1in}
\subfigure[]{
\includegraphics[width=0.45\textwidth]{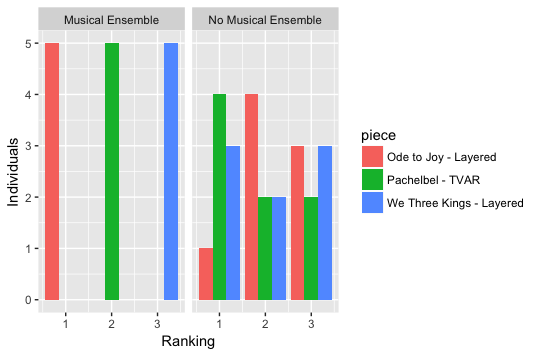}
\label{TVAR_rankings}
}
\caption{ (a) The rankings of generated pieces from a layered HMM trained on Chopin's Marche funebre and Mendelssohn's Hark! The Herald Angels Sing and the piece generated from a first order HMM trained on Beethoven's Ode to Joy as evaluated by eight listeners who were currently in a musical ensemble (left) and not currently in a musical ensemble (right).  (b) The rankings of generated pieces for the validation evaluation.}
\end{figure}

The human listeners were split into two groups of eight; group A consisted of individuals in a musical ensemble and group B consisted of individuals not in a musical ensemble. All individuals listened to the top three pieces according to the numerical metrics. In \autoref{rank_music} we plot the ranking (1-3) of the each of the pieces, with 1 as the most favorite and 3 as the least favorite.


Several members in group A thought that the pieces sounded like they could have been composed by a human, but that the composition style was typical of a ``twentieth century", ``modern", ``contemporary" or ``post-classical" composer and commented that the pieces did not sound like they had been composed in the Romantic era.  In particular, the   ``atonal chords and sustained base chords" and the ``harmonic variation and non-diatonic chords" were provided as justifications for the more modern sound of the composed pieces, though one listener noted that  this could have been due to the fact that modern composers ``are less obligated to obey the rules of tonal harmony".  Several of the listeners in group A commented on how they wished there was more phrasing or structure in the generated pieces, as there was not much ``complexity" in the pieces.  Two listeners also thought that a human performer could improve the interpretation of the generated pieces by adding some phrasing to the music itself.  Five of the eight listeners commented on the similarity of the second piece to ``Hark! The Herald Angels Sing". 

One of the listeners in group B commented that overall each piece sounded ``distinct" and suggested that this ``experimental" method of composition might be applied to ``free jazz".  Another listener thought that each piece sounded ``related" to the piece it was trained on, in particular for the piece generated by a layered HMM trained on Mendelssohn's Hark! The Herald Angels Sing, a piece  that they were familiar with.  Finally, listeners in this group also commented on how the piece was repetitive and ``forgot" what had previously occurred.

\subsection{Observations in Relation to Music Theory}

One of the observations from the listening evaluations was that the piece generated by training a layered HMM on Chopin's Marche funebre was in general the most well received, with the fewest comments complaining of un-resolved or out-of-place dissonance.  Chopin's Marche funebre is built on chords that are fifths---an open, perfect interval widely used in the music of non-Western cultures.  Thus, even when there was a passing dissonance in the piece generated by training an HMM on Chopin's Marche funebre, the dissonance could be resolved by relaxation to a pure interval, resulting in a piece that sounded less dissonant.  The majority of the dissonance in the generated piece could be resolved in this way. We believe this simplicity of the harmony in the original piece by Chopin contributed to the relative success in the generation of new pieces from HMMs trained on Chopin's Marche funebre.

In contrast, Mendelssohn's Hark! The Herald Angels Sing and Beethoven's Ode to Joy are built on major chords comprised of thirds. There is a greater ``potential" for dissonance for major chords comprised of thirds as there are multiple ways that one can obtain dissonance. Indeed, the only ways for there to be consonance is for intervals that are either: (a) in perfect unison, or (b) a third, fifth or octave in the chord.  For pieces built on thirds, the dissonance could not be resolved as easily as in the case of Chopin's Marche funebre built on open fifths, thus the generated pieces sounded much more dissonant.    

This intuition from music theory is confirmed in \autoref{simple_orig} where we display the percentage of simple harmonic intervals (harmonic intervals that are an octave or less) that are either thirds (minor or major), perfect fourths or fifths, or dissonant for both the original training piece and the generated pieces.  The third in the major or minor triad is an imperfect consonance and is not as easily resolved as perfect consonances like the fifth in the triad.  This leads to more dissonance in the generated pieces.  All of the generated pieces in \autoref{simple_orig} have a much higher percentage of dissonant simple harmonic intervals then their respective training pieces.  However, Chopin's Marche funebre is the only training piece which is primarily perfect fourths or fifths (in this case fifths) as opposed to thirds, and the generated piece has the lowest percentage of dissonant intervals of the three generated pieces.  

\begin{table}[h!]
\centering
\caption[]{\raggedright Percentage of simple harmonic intervals that are thirds, perfect fourths or fifths and dissonant for the three evaluated pieces.}
\begin{tabular}{ |p{9cm}|c|p{1.5cm}|c|  }
 \hline
 Piece & Thirds & Fourths / Fifths &  Dissonant \\
 \hline\hline
 Chopin's Marche funebre & 0.0964 & 0.4608 & 0.0060\\\hline
Chopin's Marche funebre - Layered HMM &  0.2071 & 0.2893 & 0.1286 \\\hline\hline
 Mendelssohn's Hark! The Herald Angels Sing &  0.2690 & 0.2398 & 0.0741 \\\hline
Mendelssohn's Hark! The Herald Angels Sing - Layered HMM & 0.2067 & 0.2667 &  0.2433 \\\hline\hline
Beethoven's Ode to Joy  & 0.4710 &0.2101 & 0.0145 \\\hline
 Beethoven's Ode to Joy - First Order HMM & 0.2927 & 0.2134& 0.2317 \\\hline
\end{tabular}
\label{simple_orig}
\end{table}

\begin{figure}[ht]
\centering
\subfigure[]{
\includegraphics[width = 0.4\textwidth]{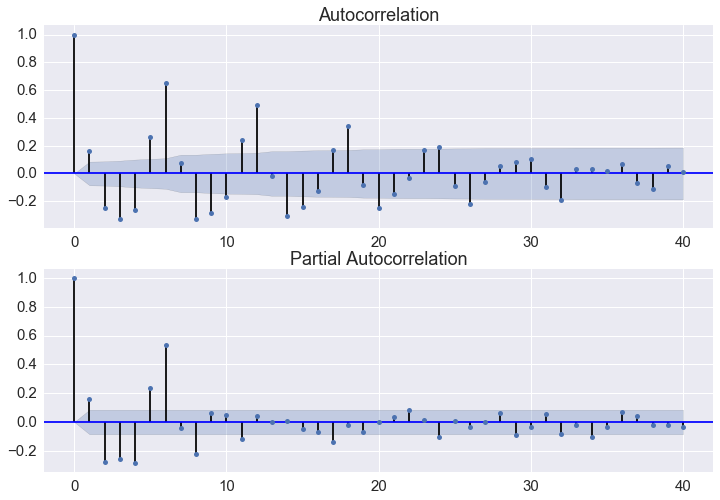}
\label{ode_acf_orig}
} \hspace{.1in}
\subfigure[]{
\includegraphics[width = 0.4\textwidth]{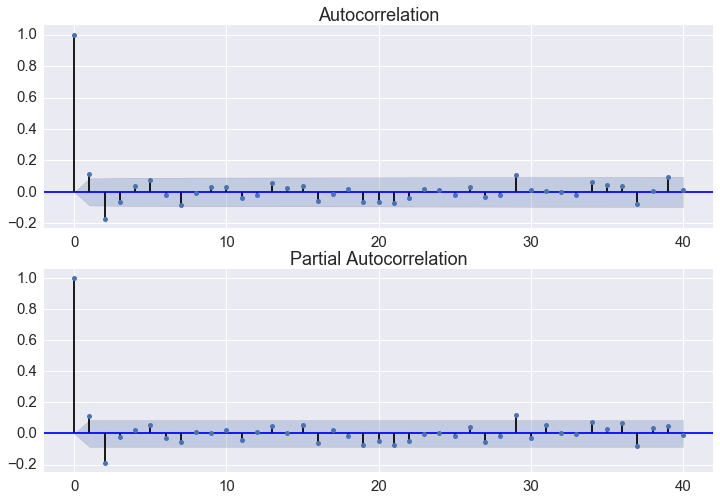}
\label{ode_acf_new}
}
\caption{ (a) Plots of the ACF and PACF for Beethoven's Ode to Joy. (b) Plots of the ACF and PACF for the piece generated by a first order HMM trained on Beethoven's Ode to Joy.}
\label{ode}
\end{figure}

All the generated pieces lacked the extent of overall melodic progression or global structure of the original pieces, a shortcoming that the listeners commented on and that was evident in the metrics. \autoref{ode} displays the ACF and PACF for both the piece generated by training a first order HMM on Beethoven's Ode to Joy and the original piece. It is clear that the original piece has much greater dependence over longer lags and even relatively short, simple, highly repetitive motifs were not modeled well by any of the HMMs considered.  This lack of global structure is to be expected, as the models considered made quite restrictive assumptions about long-term dependence between hidden states.

None of the evaluated pieces sounded like they were composed during the Romantic era.  Most listeners seemed to think that the pieces could have been composed by a human, albeit one in the modern or contemporary period (especially due to the prevalence of large intervals more than an octave and the often unresolved dissonances that are much more common in modern music than in Romantic era music).  Additionally, the layered HMM in particular, may have suffered from ``overfitting", as several listeners were able to identify that one of the evaluated pieces was indeed trained on Hark! The Herald Angels Sing.  However, we do expect the generated pieces to bear similarity to the training piece and this potential ``overfitting" is not necessarily an issue.  Many listeners could not identify the training piece, only overall themes, and none of the models reproduced the training piece exactly.

\subsection{Validation of Trends}

Two hypotheses suggested by the listener responses, the metrics and the case study were that the generated pieces lacked overall or melodic progression and that models trained on 
harmonically simple pieces tended to produce more consonant pieces that were preferred by listeners, while original pieces mainly consisting of triads led to more dissonant pieces
that listeners did not prefer.

To validate these two hypotheses we considered an additional training piece from the Baroque era, Pachelbel's Canon in D. Pachelbel's Canon in D is harmonically simple and is also built on a perfect interval, in this case the perfect fourth. Our prediction was that pieces generated from training on Pachelbel's Canon in D should result in pieces that were less dissonant than most of the ten pieces from the Romantic era.  We additionally considered the TVAR model (M13) to capture more structure over time.

Using the same procedure as the initial listening evaluations, we selected three pieces for further evaluation:
a layered HMM generated piece trained on Ode to Joy (lowest musicality RMSE),   a layered HMM piece trained on We Three Kings (lowest entropy RMSE), and a TVAR(11) generated piece trained on Pachelbel's Canon (lowest temporal RMSE). These pieces were then evaluated and ranked by a new set of human listeners, five currently in a musical ensemble (group A) and eight not currently in a musical ensemble (group B).  

Members of group A had unanimous rankings with the piece generated from Ode to Joy as their favorite and the piece generated from We Three Kings as their least favorite, see \autoref{TVAR_rankings}. The members of group B had much greater variation in preference with about half preferring the piece generated from Pachelbel's Canon as their favorite, again see \autoref{TVAR_rankings}.


One of our hypotheses was that harmonically simple pieces tended to be more consonant. To test this we examined the generated pieces from Pachelbel's Canon in more detail. Pachelbel's Canon is built upon the perfect fourth and when we consider melodic intervals, the generated piece shows a decrease in the number of dissonant melodic intervals, as compared to the original piece.  We Three Kings is also primarily built upon perfect consonances and a similar trend is observed, see  \autoref{melodic_second}. The piece generated by a layered HMM trained on Beethoven's Ode to Joy showed an increase in the percentage of dissonant melodic intervals. However, the piece generated by a layered HMM trained on Ode to Joy had the lowest percentage of dissonant melodic intervals of the three generated pieces considered, likely explaining why the harmonic aspects of this generated piece were ranked highly by both groups of listeners.  

Although the piece generated by a TVAR(11) trained on Pachelbel's Canon had the highest percentage of dissonant melodic intervals, the majority of these intervals were resolved in the piece, leading to a favorable harmonic rating by both groups of human listeners. This is in contrast to the layered HMM trained on We Three Kings which had the smallest number of dissonant harmonic intervals, though it was clear from the human listeners that the dissonant harmonic intervals that it did contain were particularly jarring. This is why
we believe that it was ranked the worst in terms of harmonic qualities.
\begin{table}[h!]
\centering
\caption[]{\raggedright Percentage of simple melodic intervals that are thirds, perfect fourths or fifths and dissonant for the evaluated validation pieces.}
\begin{tabular}{ |c|c|c|c|  }
 \hline
Piece & Thirds & Fourths / Fifths &  Dissonant \\
 \hline\hline 
Beethoven's Ode to Joy  & 0.0769 &0.3427 & 0.1119 \\\hline
 Beethoven's Ode to Joy - Layered HMM & 0.1266 & 0.2848 & 0.1709 \\\hline\hline
 We Three Kings & 0.0757 & 0.2919 &  0.4324 \\\hline
We Three Kings - Layered HMM &  0.2054 & 0.2865 & 0.1784 \\\hline\hline
Pachelbel's Canon &  0.1497 & 0.2389 & 0.5191 \\\hline
Pachelbel's Canon - TVAR(11) & 0.625 & 0.275 & 0.3000 \\\hline
\end{tabular}
\label{melodic_second}
\end{table}

\begin{figure}
\subfigure[\label{pachelbel_ACF}]
  {\includegraphics[width=.3\linewidth]{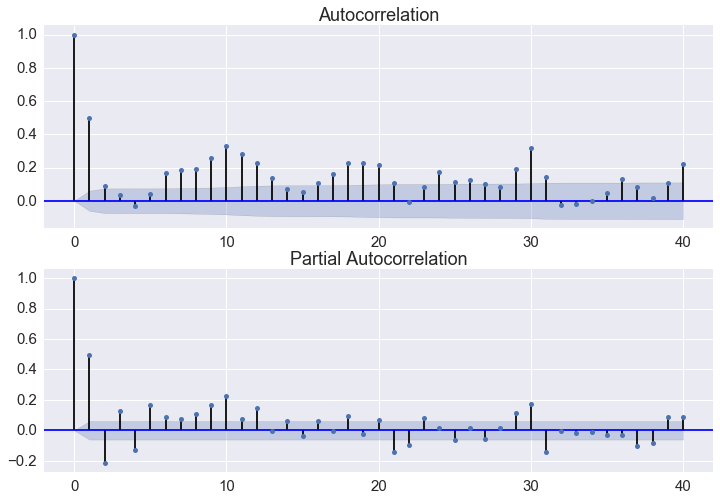}}\hfill
\subfigure[\label{layered_ACF}]
  {\includegraphics[width=.3\linewidth]{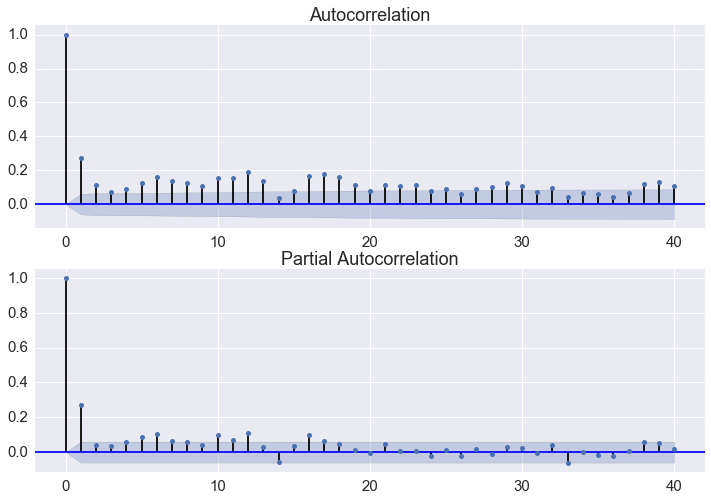}}\hfill
\subfigure[\label{TVAR_ACF}]
  {\includegraphics[width=.3\linewidth]{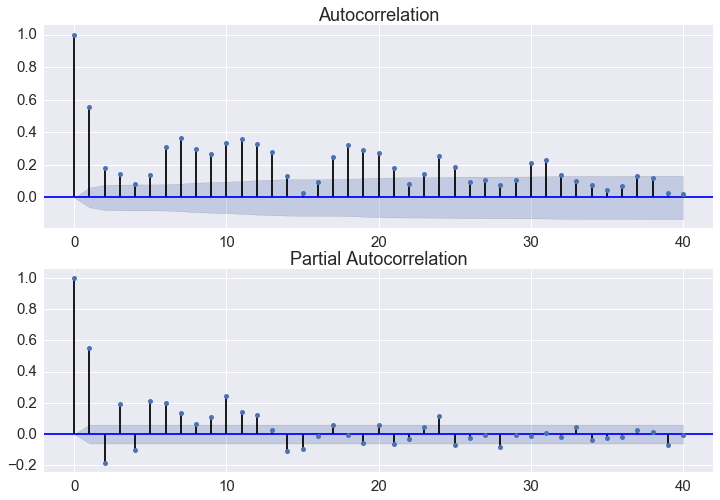}}
\caption{(a) ACF and PACF for the original training piece, Pachelbel's Canon. (b) ACF and PACF for a piece generated by a layered HMM trained on Pachelbel's Canon. (c) ACF and PACF for a piece generated by a TVAR(11) model trained on Pachelbel's Canon. }\label{pachelbel}
\end{figure}

Our other hypothesis was that the generated pieces lacked overall or melodic progression as compared to the originals. We explored this hypothesis by examining if the TVAR model, which is capable of modeling longer lags than the the layered HMMs, improved the musicality metrics. In \autoref{TVAR_ACF} we see that the decay in correlations is not as steep for the TVAR model as it is for the  layered HMM, though the autocorrelation function for both models is impoverished with respect to the original. However, human listeners were not always able to distinguish this additional temporal structure, as the TVAR(11) generated piece trained on Pachelbel's Canon was not consistently scored higher for melodic qualities than the pieces generated by a layered HMM.

\section{Conclusions and Future Work}\label{sec:conclusions}

We were able to train state space models on pieces from the Romantic era and some of the generated pieces were considered to be possibly composed by humans when heard by human evaluators. The models were more successful at modeling harmony than melodic progression and the main aspects of harmony were successfully encoded in the state space. The state space models were more successful at generating consonant pieces when trained on pieces with simple harmonies, particularly original pieces that were built on perfect intervals.  Models with greater hierarchical structure, particularly the layered HMM, tended to be more successful at generating pieces with less dissonance and slightly more melodic progression than other models considered.
Listeners felt that the generated pieces  sounded more like pieces composed in the Modern era than like pieces composed in  the Romantic era. The greatest shortcoming was that the generated pieces lacked global structure or long-term melodic progression. 

Based on the shortcomings of the state space models, several directions for future work are suggested that attempt to resolve some of the problems with the generated pieces of the considered models. These future directions include considering hierarchical models, natural language models, and recurrent neural networks (RNNs, \citep{Goodfellow:EtAl:2016}).

Music contains several layers that evolve over different time periods and of the models we considered, those that had a hierarchical component tended to perform well.  Thus, models with  a greater hierarchical structure are promising to explore in the future.  Another area for future work is exploring natural language models for music to improve the harmonic and melodic aspects of the generated pieces. Finally, one of the main criticisms of the music composed by the state space models we considered was the lack of global structure in the generated pieces.  Models that are capable of modeling longer term structure, such as RNNs  and Long Short Term Memory units  have found success in areas related to music composition \citep{Mozer:1994,Eck:Schmidhuber:2002,Boulanger-Lewandowski:EtAl:2012,Johnson:2015,MagentaDevSummit:2017,Magenta,Hadjeres:Pachet:2016} and are promising candidates for improving the global structure of the generated musical pieces.


\vspace*{2cm}\vspace*{2cm}\vspace*{2cm}\vspace*{2cm}\vspace*{2cm}

 \section*{Software and Data}

\noindent All the software to generate the music, compute the metrics, as well as the original pieces and the
generated pieces are available at \url{https://aky4wn.github.io/Classical-Music-CompositionUsing-State-Space-Models/}.
In addition examples of music theory concepts are provided and
discussed.

\section*{Acknowledgements}
A.Y.~would like to acknowledge Jeff Miller for discussions concerning the Two Hidden State HMM, Mike West for providing TVAR code in Matlab, which was converted to Python code, and Mike Kris for many discussions about the musical aspects of this work.  S.M. would like to acknowledge NSF DMS 16-13261, NSF IIS 15-46331, NSF DMS 14-18261,   NSF DMS-17-13012, NIH R21 AG055777-01A, and NSF ABI 16-61386 for partial support.


\clearpage
\newpage
\bibliographystyle{chicago}
\bibliography{JASArefs}

\clearpage
\newpage

\section*{Supplementary Materials}
\setcounter {section} {0}
\section{Time Series Model Likelihoods}

The time series models considered in this paper are all variations of HMMs \citep{Rabiner:1989} with the observable variables  $\{X_1,...,X_T\}$ 
given by our pitch representation. We will denote an observed sequence $\{x_1,...,x_T\}$ as $x_{1:T}$. The likelihood for the basic HMM is
\begin{equation}
p(x_{1:T} \mid z_{1:T}, \theta,\theta',\pi) = \pi(z_1) \, p_{\theta}(x_1 \mid z_1) \, \prod_{t=2}^T p_{\theta'}(z_t \mid z_{t-1}) \, p_\theta(x_t \mid z_t),
\end{equation} 
where $z_{1:T}$ are the hidden states and the model parameters $\theta$ specify the emission probabilities $p_\theta(x \mid z)$, $\theta'$ specify the transition probabilities $p_{\theta'}(z_t \mid z_{t-1})$, and the initial distribution is $\pi$. The standard approach to infer model parameters is the Baum-Welch Algorithm \citep{Baum:1972}.

In addition to the standard HMM, we considered a variety of statistically richer Markov models. More complicated and longer range dependence structures were modeled using
higher-order Markov models (k-HMMs), where k specifies the order of dependence, and  Autoregressive HMMs (ARHMMs). Time varying transition probabilities were modeled 
via Left-Right HMMs  (LRHMMs) and hidden semi-Markov models (HSMMs). More complex hierarchical state structure was modeled using
two state hidden Markov models (TSHMMs), factorial HMMs (FHMMs), and Layered HMMs (LHMMs).  Time-Varying Autoregression (TVAR) models were explored to model the  non-stationarity and  periodicity of the note pitches for the original pieces. We provide a brief summary of each of the models below.

{\em k-HMMs}  specify the state transition probabilities based on the present and past (k-1)-states rather than just the current state.  For example, the likelihood of a 2-HMM can be specified as 
\begin{displaymath}
p(x_{1:T} \mid z_{1:T}) = \pi(z_1) \, p(x_1 \mid z_1) \, p(z_2 \mid z_1) \, p(x_2 \mid z_2) \, \\ \prod_{t=3}^T p(z_t \mid z_{t-1}, z_{t-2}) \, p(x_t \mid z_t).
\end{displaymath} 

The Baum-Welch algorithm for parameter inference can be easily extended to these higher order models \citep{Mari:Schott:2001}.

{\em ARHMMs} are extremely common in time series applications \citep{Rabiner:1989,Guan:EtAl:2016} and in our setting allow each note to depend not just on the hidden state at time $t$ but also on the previous observed note . The likelihood for the ARHMM model is
$$p(x_{1:T} \mid z_{1:T})  =  \pi(z_1)  p(x_1 \mid z_1)  \left[ \prod_{t=2}^T p(z_t \mid z_{t-1}) p(x_t \mid z_t, x_{t-1}) \right].$$

{\em LRHMMs} adapt standard HMMs by constraining the state transition matrix to be upper triangular. This constraint allows for time varying transitions as the chain starts at an initial state and  traverses a set of intermediate states and terminates at a final state without being able to go ``backwards."

{\em HSMMs} explicitly model time varying transitions by modeling the duration for each hidden state. This relaxes the stationarity condition of standard HMMs and specifies a semi-Markov chain \citep{Yu:2010,Murphy:2002}. The motivation for HSMMs in music is that we might expect for a piece to remain in the same state over the course of a few bars, for example over a motif, and the HSMM allows us to model this variable state duration.  The likelihood for a hidden semi-Markov model can be specified as
$$p(x_{1:T'} \mid z_{1:T},d_{1:T})  =  \pi(z_1) \prod_{j=1}^{d_1} p(x_j \mid z_1)  \left[ \prod_{t=2}^{T} p(z_t \mid z_{t-1}) \prod_{j=1}^{d_t} p(x_{{j_t}+j} \mid z_t) \right],$$
here $d_t$ is a random count that determines the duration of state $z_t$, $T' = \sum_{t=1}^T d_t$ is the length of the observed sequence and $j_t$ is the index of the first observation generated from hidden state $z_t$ with $j_t = \sum_{j=1}^{t} d_j$ for $t>1$. Note in this model the sequence generating process $d_{1:T}$ also needs to be specified and its parameters need to be inferred.

{\em FHMMs} allow for a distributed representation for the hidden states \citep{Ghahramani:Jordan:1997}.  An example of a factorial HMM is $m$ independent Markov chains to update the hidden states with the  observation distribution specified by a function of the  $m$ hidden states. The FHMM model we specified is
\begin{eqnarray*}
p(x_{1:T} \mid z_{1,1:1,T},..., z_{m,1:m,T}) &= & \prod_{j=1}^m \pi(z_{m,1}) \, p(x_1 \mid z_{1,1},...,z_{m,1}) \times \\
& &  \prod_{t=2}^T \left[  \prod_{j=1}^m p(z_{m,t} \mid z_{m,t-1}) \right] p(x_t \mid z_{1,t},...,z_{m,t}),
\end{eqnarray*}
here $z_{j,1:j,T}$  is the sequence of the hidden states for the $j$-th chain and the observation process is based on the average of the states of the $m$
hidden states
$$p(x_t \mid z_{1,t},...,z_{m,t}) = p\left(x_t \mid \frac{1}{m}  \sum_{j=1}^m  z_{j,t}\right).$$
In our case $m=3$ and the number of states in each chain may vary. In our application the FHMM allows us to  model different dynamic processes independently and combine them 
to form a generated piece of music.

{\em LHMMs} are complex models based on stacking HMMs and at each layer a standard HMM is learned on top of the previous layers \citep{Oliver:EtAl:2004}.  
The lowest level is a standard HMM. Once this HMM is trained the most likely sequence of hidden states is calculated and then used as observations for the HMM in the next layer.
We considered three layers of HMMs, all with the same number of possible hidden states.

{\em TVAR} models \citep{Prado:West:2010} are able to model non-stationarity and longer dependence between note pitches over time.  From both a musical and modeling perspective, we expect the note pitches for each piece to be volatile over time. The TVAR model for a sequence $(X_1,...,X_T)$ is specified as
$$X_{t} = \sum_{j=1}^d \theta_j ((t-1)/T) X_{k-j} + \sigma(t/T)  \varepsilon_t, \quad t=1,...,T$$
where the variance $\sigma(t/n)$ is time varying,  $\varepsilon_t$ can also be time varying, and  the parameters $\theta_j ((t-1)/T)$ are also time varying.
We used forward filtering to fit each TVAR model and backwards sampling \citep{Prado:West:2010} was used to generate new pieces. The order and discount factors of the TVAR models
were selected by a grid-search to maximize the  marginal likelihood. The orders of the TVAR models ranged from 7 to 14, depending on the training piece. Since the note pitch values were discrete (rather than continuous) and only certain note pitches occurred in the original piece, the generated note pitches were binned to the nearest pitch that occurred in the original piece.

{\em TSHMMs} impose a hierarchical structure on the hidden states by considering  two hidden states $r,s$ where $s_t$ is conditionally dependent on both its previous state $s_{t-1}$ and
the current state of $r$, while $r_t$ is only dependent on its previous state $r_{t-1}$. The likelihood for this model is 
\begin{displaymath}
p(x_{1:T} \mid s_{1:T},r_{1:T}) = \pi(s_1) \, p(x_1 \mid s_1) \, \prod_{t=2}^T (r_t \mid r_{t-1}) \, p(s_t \mid r_t, s_{t-1}) \ p(x_t \mid s_t).
\end{displaymath} 
A derivation of the Baum-Welch Algorithm for parameter inference for this model is given below.

\begin{figure}[tbp]
\begin{center}
\includegraphics[width=75mm]{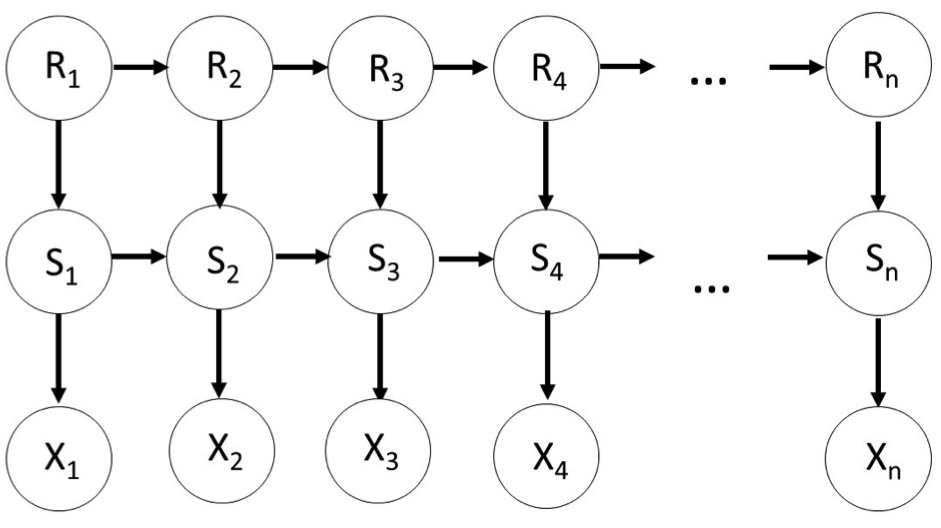}
\end{center}
\caption{Graphical Model for the TSHMM.}
\label{twinkle}
\end{figure}

$R_{1:n}$ and $S_{1:n}$ are the hidden states.  Each state in the hidden process $S_{1:n}$ can take on one of $m_1$ possible values, while each state in the hidden process  $R_{1:n}$ can take on one of $m_2$ possible values.  The length of both series is still $n$.  We define the following parameters:
\begin{equation*}
\begin{split}
C_{ij} &= P(R_t = j | R_{t-1} = i)\\
D_{j,k,l} &= P(S_t = l | R_t = j, S_{t-1} = k)\\
A_{ik, jl} &= C_{ij} D_{jkl} = P(R_t = j, S_t = l | R_{t-1} = i, S_{t-1} = k)\\
Z_t &= (R_t, S_t)
\end{split}
\end{equation*}

\noindent The constraints are $\sum_j C_{ij} = 1, \sum_l D_{jkl} = 1.$ 

Let $\theta = (\pi, A, B, C, D)$ be the model parameters, where $\pi$ and $B$ are the initial state distribution and emission distribution, respectively, as defined for the first order HMM.  Let $\theta^{(t)}$ be the current values of these parameters at time $t$ in the Baum-Welch Algorithm.  Define $c$ to be a constant. Then,  the auxiliary function for the E step of the update Baum-Welch Algorithm for the HMM with two hidden states  can be written as: \newline
\begin{equation*}
\begin{split}
Q(\theta, \theta^{(t)}) &= \mathbb{E}_{\theta^{(t)}}(\log p_{\theta} (X_{1:n}, Z_{1:n} | X_{1:n} = x_{1:n})) \\
&= c + \sum_{t=2}^n \sum_{i,k} \sum_{j,l} P_{\theta_k} (R_{t-1} = i, S_{t-1} = k, R_t = j, S_t = l | X_{1:n}) \log C_{ik, jl} \\
\end{split}
\end{equation*}
Let $$D_{t,ik,jl} = P_{\theta^{(t)}} (R_{t-1} = i, S_{t-1} = k, R_t = j, S_t = l | X_{1:n}).$$ We have that $$\log A_{ik, jl} = \log C_{ij} + \log D_{j,k,l}.$$

Then, we can find the value of $\theta$ to maximize $Q(\theta, \theta^{(t)})$ (where $\nu$ is a Lagrange multiplier to handle the constraints placed on $C$ and $D$): \newline
\begin{equation*}
\begin{split}
 0 &= \frac{\partial}{\partial C_{ij}}\left(Q(\theta, \theta^{(t)}) - \nu \sum_{j} C_{ij}\right) \\
0 &= \left(\sum_{t=2}^n \sum_k \sum_l D_{t,ik,jl} \frac{1}{C_{ij}}\right) - \nu \\
\nu C_{ij} &= \sum_{t=2}^n \sum_k \sum_l D_{t,ik,jl} \\
\nu &= \sum_j \sum_{t=2}^n \sum_k \sum_l D_{t,ik,jl} \\
C_{ij} &\propto \sum_{t=2}^n \sum_{k,l} D_{t,ik,jl} \quad \forall 1\leq i,j \leq m_2 \\
\end{split}
\end{equation*}

Likewise,
\begin{equation*}
\begin{split}
0 &= \frac{\partial}{\partial D_{j,k,l}} \left(Q(\theta, \theta^{(t)}) - \nu \sum_l D_{j,k,l}\right) \\
&= \sum_{t=2}^n \sum_i D_{t,ik,jl} \frac{1}{D_{j,k,l}} - \nu \\
 \nu D_{j,k,l} &= \sum_{t=2}^n \sum_i D_{t,ik,jl} \\
D_{j,k,l} &\propto \sum_{t=2}^n \sum_i D_{t,ik,jl} \quad \forall 1\leq l,k\leq m_1, 1\leq j \leq m_2 \\
\end{split}
\end{equation*}

The Forward-Backward Algorithm is exactly the same as in the first order HMM case, where $A$ as defined above is the transition matrix used.  $\pi$ and $B$ are updated exactly the same way as in the Baum-Welch Algorithm for the first order HMM (\cite{Miller:2016b}).

\section{Music Theory}
For this work, we consider a high-level view of several aspects of music theory, while  \cite{Laitz:2003} and \cite{Gauldin:2004} provide more in-depth treatments of musical theory.

\paragraph{Grand Staff and Note Pitches}
The grand staff is used for most piano music and is composed of the lower Bass Clef part (usually played with the left hand when playing the piano) and the higher Treble Clef part (usually played with the right hand). Each subsequent key on a piano corresponds to a half step in note pitch. The white keys correspond to the natural notes, while the black keys raise or lower the neighboring white key pitches by a half step, resulting in a sharp or flat note.

\begin{figure}[tbp]
\begin{center}
\includegraphics[width=\textwidth]{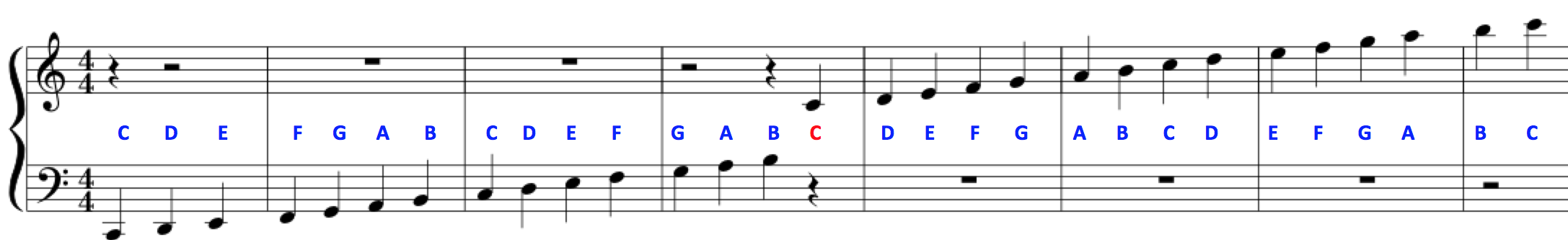}
\end{center}
\caption{Four octaves of the C-Major scale with labeled note pitches, where each note is one quarter note in duration. The note pitch highlighted in red corresponds to middle C.}
\label{grand_staff}
\end{figure}

The numeric markings next to the clef in the grand staff indicate the time signature of the piece. The top number indicates how many beats per measure there are in the piece and the bottom note indicates which note gets the beat.  For the example in \autoref{grand_staff}, the time signature is ``four-four”, meaning that there are four beats per measure and the quarter note gets the beat.

The key signature indicates the key that the piece is to be played in. For the example in \autoref{grand_staff}, there are no sharps or flats in the key signature, so the key of this example is C-Major. There can be both major and minor keys. This example contains four octaves of the C-Major scale, where each note is one quarter note in duration. 

\paragraph{Intervals}
There are two types of intervals that occur in music, harmonic intervals where two or more notes are played at the same time, and melodic intervals, where two or more notes are played sequentially. Most melodic and harmonic intervals occurring in the Romantic era were smaller than an octave. 

\paragraph{Chords}
A chord is a combination of three or more different pitches, and the predominant chords occurring in the Romantic era were the triad and the seventh chord. The triad is composed of a root pitch and the third and fifth above the root, while a seventh chord is the root, third, fifth and seventh. Chords can be closed or open depending on the spacing between pitches in the higher and lower registers.

\begin{figure}[tbp]
\begin{center}
\includegraphics[width=150mm]{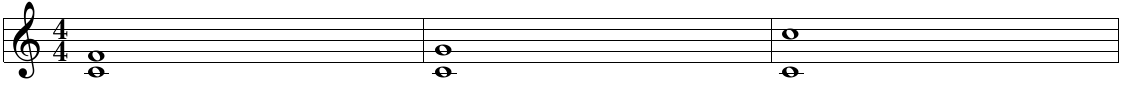}
\end{center}
\caption{Perfect fourth, perfect fifth and perfect octave harmonic intervals in C-Major.}
\label{perfect}
\end{figure}

\paragraph{Dissonance and Consonance}
Some simple (less than an octave) intervals are considered consonant and do not need to be resolved, while other intervals are dissonant and require a resolution. The octave, perfect unison and the third, fifth and sixth intervals are consonant intervals, while the second and seventh intervals are considered dissonant. The perfect fourth is considered dissonant in some contexts and consonant in others. The perfect unison, octave, perfect fourth and perfect fifth are perfect consonances (\autoref{perfect}), in that they are stable intervals that do not need to resolve to a more stable pitch. Thirds and sixths are imperfect consonances and are moderately stable; they again do not need a resolution. Dissonant intervals on the other hand need to be resolved to consonant intervals. The tritone, also known as an augmented fourth or a diminished fifth, is considered dissonant.

\begin{figure}[tbp]
\begin{center}
\includegraphics[width=\textwidth]{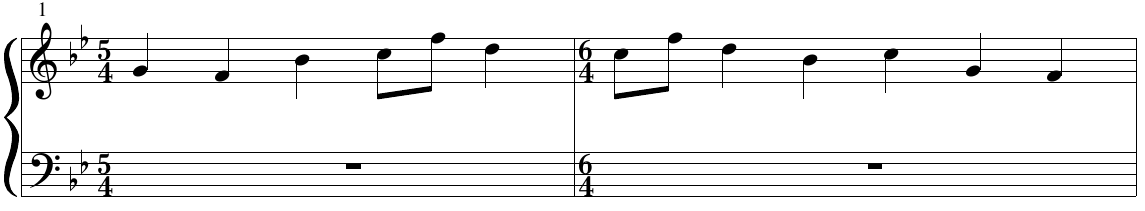}
\end{center}
\caption{Example of a musical motif: Mussorgsky’s Pictures at an Exhibition - Gnomus.}
\label{muss}
\end{figure}

\paragraph{Motifs}
A motif is a melodic or rhythmic unit that reappears throughout a piece and is shorter than a theme. The combination of phrases and motifs forms a melody in a piece. A motif can reappear in its original form or at different pitches or with different intervals throughout a piece. An example motif is the opening two bars of Mussorgsky’s Pictures at an Exhibition - Gnomus (\autoref{muss}).

\paragraph{Romantic Era Music}
Romantic era music is considered tonal music, meaning that the music is oriented around and pulled towards the tonic pitch. The Romantic period of music lasted from the end of the eighteenth century to the beginning of the twentieth century. This period was marked by a large amount of change in politics, economics and society. In general, music from the Romantic era tended to value emotion, novelty, technical skill and the sharing of ideas between the arts and other disciplines. Some key themes of Romanticism were the emphasis of emotion over reason, national pride, the extolling of common people and an interest in the exotic. Romantics emphasized the individual, national identity, freedom and nature and the Romantic era saw an increase in virtuosity in musical performance \citep{Warrack:1983}.

The piano as an instrument also changed considerably during the Romantic era, resulting in new trends in composition for the instrument. Technical and structural improvements to the piano lead to an instrument that was capable of a larger range of pitches and dynamics as compared to the pianoforte of the seventeenth and early eighteenth centuries. Improvements in the sustaining pedal in particular enabled the piano to create a more dramatic and sustained sound. Furthermore, the piano saw an increased presence in musical, commercial and social circles in the Romantic era \citep{Ratner:1992}.

\section{Case Study: Twinkle, Twinkle, Little Star}

In order to explore the learned model parameters in terms of music theory, we considered the simple tune ``Twinkle, Twinkle, Little Star" as a case study.  Twinkle, Twinkle, Little Star was selected for it's simplicity in terms of melody, harmony and rhythm, allowing for an easier evaluation of model parameters.

We considered a version of the piece in the key of C major (\autoref{twinkle}).  Twinkle, Twinkle, Little Star is a tonal piece, with the main harmonies consisting of the major triad and intervals of the perfect fourth and the perfect fifth.  The harmony primarily consisted of perfect consonances and there were no dissonant harmonic intervals in the original piece.  These harmonies did not change over the course of the piece and the harmonies were thus stationary.  The melody in the higher notes was supported by the harmony in the base notes.  The melody consisted of two bar phrases, first an ascending phrase for two measures, followed by a descending two bar phrase.  There were two subsequent two bar descending phrases that mirrored the first, then the piece ended with the ascending and descending phrase from the beginning of the piece (\autoref{twinkle}).

\begin{figure}[tbp]
\begin{center}
\includegraphics[width = \textwidth]{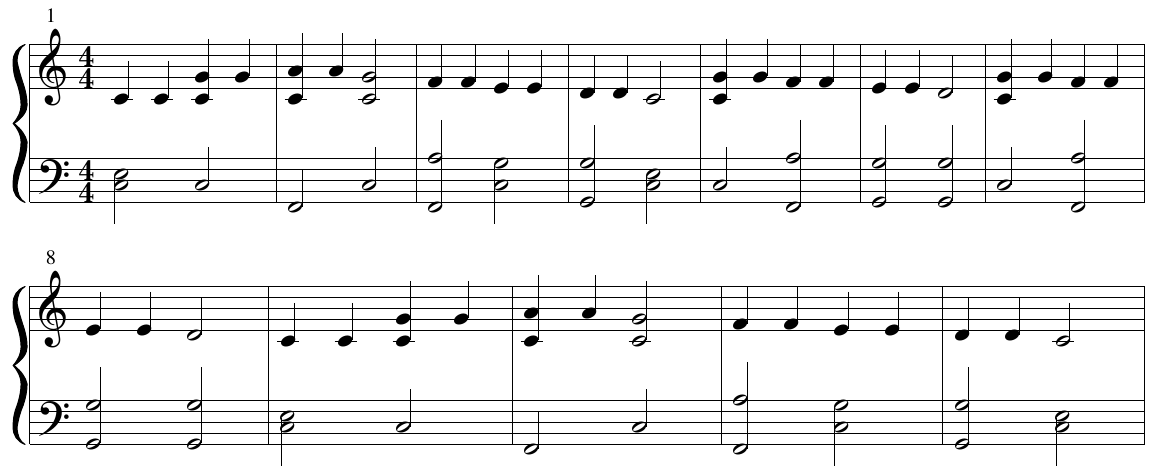}
\end{center}
\caption{Sheet music for Twinkle, Twinkle, Little Star.}
\label{twinkle}
\end{figure}

Twinkle, Twinkle, Little Star is a tonal piece, with the main harmonies consisting of the major triad and intervals of the perfect fourth and the perfect fifth.  While there were some imperfect consonances in the piece, the harmony primarily consisted of perfect consonances and there were no dissonant harmonic intervals in the original piece.

\begin{figure}[tbp]
\begin{center}
\includegraphics[width=\textwidth]{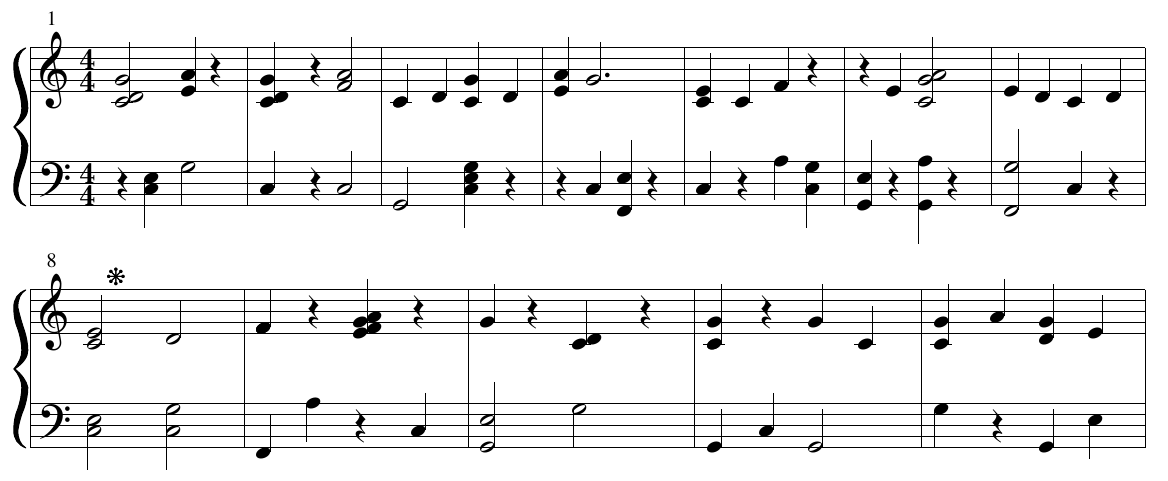}
\end{center}
\caption{Sheet music for a piece generated by a HMM with 5 hidden states.}
\label{twinkle_first}
\end{figure}

\paragraph{Additional Hidden States}
As the number of hidden states was increased, the trends seen from five hidden states (\autoref{twinkle_first}) held.  There was always at least one hidden state that corresponded to each primary harmony in the original piece and the initial distribution had highest probability on hidden states that corresponded to the tonic pitch. The transition matrices also had fairly equal probability in transitioning to each hidden state.  However, even as the number of hidden states was increased, there was never a hidden state that had highest emission probability for A (the major sixth) or D (the major second).  These ``secondary" harmonies that occurred less frequently were not captured by the first order HMM, even when the number of hidden states was increased.

 \begin{table}[h!]
 \small
\caption[]{\raggedright\small Evaluation metrics for each of the models considered for Twinkle, Twinkle, Little Star for 1000 generated pieces.  Ent. corresponds to the average empirical entropy of the generated pieces, while MI and ED correspond to the mutual information and edit distance, respectively, calculated with respect to the original piece and averaged over all the generated pieces.  The remaining metrics are the RMSE between the original piece and the generated pieces, averaged over all metrics in that category.  For example, ACF is calculated out to lag-40 and the value reported is average RMSE value over all 40 lags.  Per. refers to the percentage of intervals (both harmonic and melodic) that are perfect consonances, imperfect consonances and dissonances.  H. Ints refers to harmonic intervals and M. Ints refers to melodic intervals. NC refers to the note counts in the piece.  The best performing model for each metrics is highlighted in blue and the worst in red.}
\centering
\begin{tabular}{|c|c|c|c|c|c|c|c|c|c|} 
\hline
Model &	Ent. &	MI &	ED &	H. Ints & 	M. Ints &	Per. & 	NC & 	ACF & 	PACF\\ [0.5ex] 
 \hline\hline
HMM	&2.315	&0.400	&0.742	&\textcolor{blue}{11.360}	&\textcolor{blue}{11.738}	&0.091	&0.033	&0.244	&0.222\\\hline
LRHMM	&2.371	&0.386	&0.765	&11.673	&12.662	&0.098	&0.022	&0.245	&0.223\\\hline
2-HMM	&2.339	&0.383	&0.777	&12.578	&13.358	&0.096	&0.026	&0.245	&0.222\\\hline
2-LRHMM	 &2.384	&0.382	&0.780	&12.594	&13.905	&0.101	&0.024	&0.245	&0.223\\\hline
3-HMM	&2.397	&0.386	&0.780	&12.448	&13.662	&0.100	&0.021	&0.246	&0.223\\\hline
3-LRHMM	 &2.377	&0.380	&0.784	&12.616	&13.504	&0.098	&0.023	&0.246	&0.223\\\hline
Random	& \textcolor{blue}{2.426}	&0.389	&0.812	&13.024	&\textcolor{red}{13.947}	&0.103	&0.040	&0.246	&\textcolor{red}{0.224}\\\hline
HSMM	&2.324	&0.386	&0.767	&11.740	&12.600	&0.095	&0.030	&\textcolor{red}{0.246}	&0.223\\\hline
ARHMM	&2.374	&0.382	&0.786	&12.565	&13.560	&0.098	&0.020	&0.245	&0.223\\\hline
TSHMM 5-2	&1.896	&0.284	&0.773	&9.224	&9.807	&0.118	&0.075	&0.225	&0.211\\\hline
TSHMM 2-5	&\textcolor{red}{0.829}	&\textcolor{blue}{0.068}	&\textcolor{blue}{0.872}	&8.020	&13.868	&\textcolor{red}{0.126}	&\textcolor{red}{0.117}	&0.245	&0.223\\\hline
FHMM	&2.216	&0.754	&0.750	&13.030	&13.277	&0.106	&0.051	&\textcolor{blue}{0.146}	&\textcolor{blue}{0.173}\\\hline
LHMM	&2.381	&\textcolor{red}{1.143}	&\textcolor{red}{0.607}	&11.838	&12.667	&\textcolor{blue}{0.089}	&\textcolor{blue}{0.018}	&0.218	&0.212\\\hline
TVAR	&2.287	&0.509	&0.822	&\textcolor{red}{13.327}	&13.577	&0.106	&0.050	&0.203	&0.184\\\hline
\end{tabular}
\label{table:twinkle_metrics}
\end{table}

\paragraph{Random HMM}
As the random HMM did not learn from the observed notes, all of the parameters looked quite different from the parameters learned by the actual hidden Markov models.  For each hidden state, the random HMM had emission probabilities for all notes that were much closer to each other than for the learned parameters.  Furthermore, since the random HMM had no relation to the observed notes, apart from generating pieces with the highest average entropy, the random HMM performed poorly on all metrics.

\paragraph{Second and Third Order HMMs}
Both the 2-HMM and 3-HMM emission distributions showed similar trends of having hidden states represent the main harmonies in the piece.  However, since the HMM was already able to represent the changes between these main harmonies fairly well in the transition matrix, the 2-HMM and 3-HMM with the extra dependence among the hidden states did not encode more information about the harmonic progression of the musical piece than the HMM.  Additionally, since the harmonies in Twinkle,Twinkle, Little Star primarily progressed based on the previous harmony, the autoregressive dependence in the hidden states did not capture more harmonic aspects of the piece and the 2-HMM and 3-HMM did not outperform the HMM in terms of any of the metrics (\autoref{table:twinkle_metrics}).  

The hidden states further seemed to represent the main harmonies and not aspects of melody, as the 2-HMM and 3-HMM did not outperform the first order HMM in the temporal metrics.  The hidden states were thus not encoding very much about the melody of the original piece.

\paragraph{Left-Right Models}
The left-right models placed explicit restrictions on the form of the transition matrix.  Given that the hidden states tended to represent the main harmonies in the piece and that these main harmonies could transition among each other in terms of music theory, the left-right restriction on the transition matrix only served to increase the probabilities in the transition matrix so that there were less transitions between hidden states.  This did not improve the harmony or the melody of the generated piece, and thus the HMM tended to outperform the higher order LR models (as above) and the LR-HMM on the majority of the metrics considered.  The LR models  also failed to capture all of the main harmonies in the original, for example, the LRHMM did not have a hidden state that tended to emit the major triad or perfect fourth.

\paragraph{HSMM}
The HSMM, which explicitly allowed for multiple time steps in the same hidden state performed comparably to the HMM and LR-HMM in terms of the harmonic and melodic interval metrics, as the HMM was already able to adequately model the primary harmonies and transition among them.  Thus, the HSMM did not offer any explicit harmonic improvements over the HMM.  The HSMM performed poorly on the temporal metrics, perhaps because the restriction on the dynamics of the harmonies further restricted the ability of the model to capture melody in the piece.

\paragraph{ARHMM}
The ARHMM added additional dependence in the observed states.  The hidden states were again able to capture the predominant harmonies for the ARHMM.  This 1-lag autoregressive structure in the observed notes was not enough to capture more melodic or temporal structure in the original piece.  Even in this simple piece, the melody depended on more than just the previous note.  

\paragraph{TSHMM}
The TSHMM was examined with two configurations, one with 5 hidden states for the lower level and 2 hidden states for the second level and vice versa. This hierarchical model was expected to capture additional dynamics in the system that could be moving at different rates.  In practice, however, the additional layer of hidden states did not capture any more about the harmonies in the piece beyond the HMM.  While some of the secondary harmonies had higher emission distribution probabilities than in the HMM, these secondary harmonies still had lower emission probabilities than the main harmonies and were thus not captured by the additional layer of hidden states.

\begin{figure}[tbp]
\begin{center}
\includegraphics[width=\textwidth]{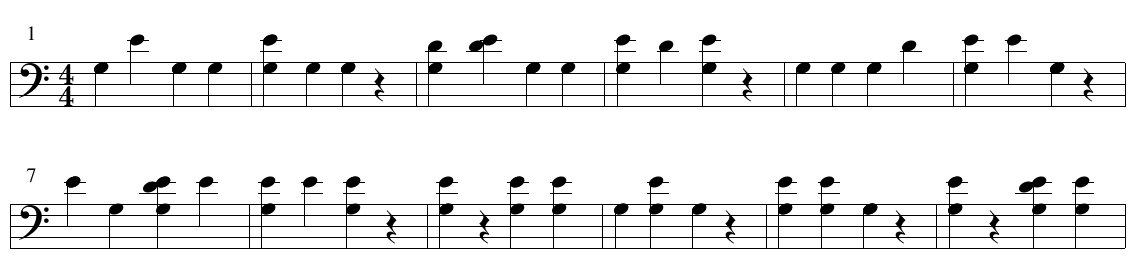}
\end{center}
\caption{Sheet music for a piece generated by a TSHMM with 2 hidden states in the first layer and 5 in the second layer.}
\label{twinkle_TSHMM2}
\end{figure}

The TSHMMs tended to produce pieces with the lowest average entropy and performed poorly in terms of the percentage of consonant harmonic and melodic intervals.  The additional layer of hidden states appeared to be hindering the ability of the harmonies to be modeled, perhaps by trying to split the hidden states into two aspects when there was only one layer needed to capture the main harmonies of the piece.  The generated pieces tended to be very repetitive and have little to no harmony  or chords (\autoref{twinkle_TSHMM2}).  Furthermore, the tonic note was not observed in the TSHMM generated piece.   

\paragraph{Factorial HMM}
The  FHMM considered was the average of three HMMs.  While the  FHMM tended to perform poorly on the harmonic and melodic interval metrics, it outperformed the HMM in terms of the temporal metrics.  The averaging of the parameters hindered the ability of the model to capture harmony, as each hidden state for each HMM represented a specific predominant harmony, but the order of these hidden states was exchangeable, so when the emission distributions were averaged, the result was a worse modeling of the harmonies in the original piece and dissonant generated pieces.  

However, by averaging three HMMs, the  FHMM did seem to be able to capture more about the melodic and temporal aspects of the piece.  A  FHMM with dependencies between each HMM would likely alleviate the harmonic problems with the independent  FHMM, but the combination of several individual HMMs seemed able to capture more melody than a HMM alone. 

\begin{figure}[tbp]
\begin{center}
\includegraphics[width=\textwidth]{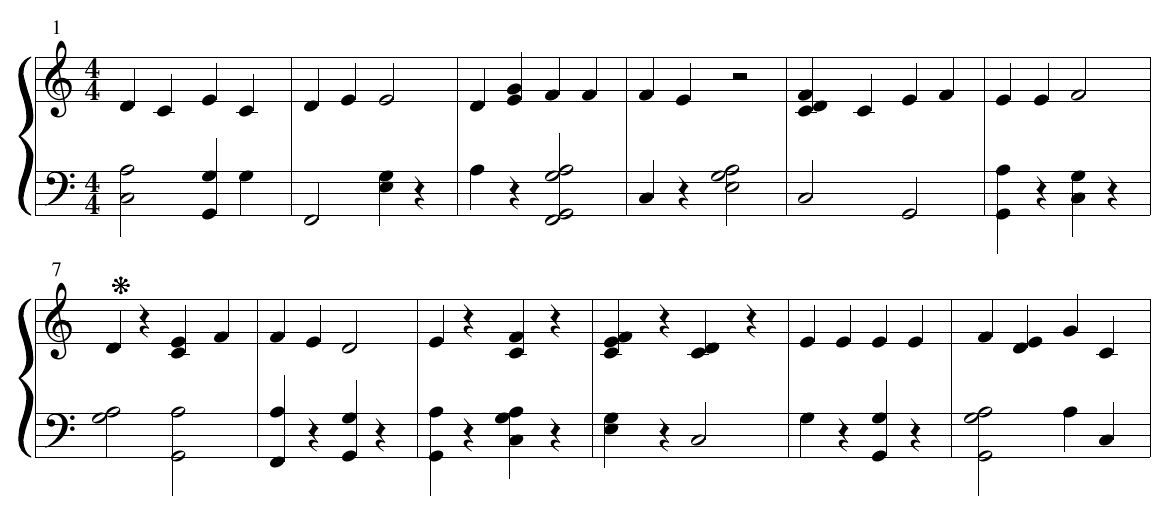}
\end{center}
\caption{Sheet music for a piece generated by a FHMM with 5 hidden states in each of three independent HMMs.}
\label{twinkle_factorial}
\end{figure} 

\begin{figure}[tbp]
\begin{center}
\includegraphics[width=\textwidth]{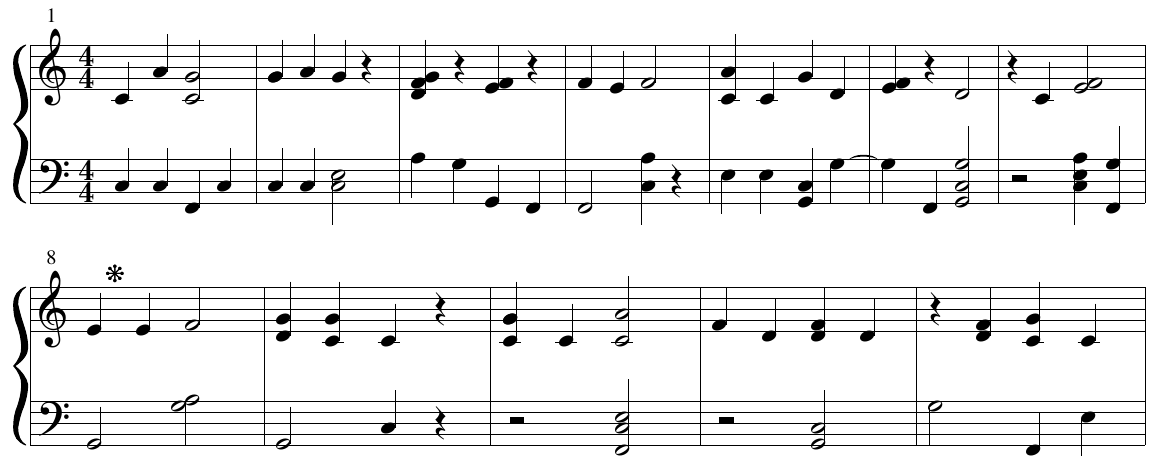}
\end{center}
\caption{Sheet music for a piece generated by a LHMM with 5 hidden states.}
\label{twinkle_layered}
\end{figure}

\paragraph{Layered HMM}
The LHMM had the highest average mutual information with the original piece and the lowest average edit distance, indicating that the LHMM tended to produce pieces that were the most similar to the original piece.  However, this also meant that the LHMM performed well in terms of the harmonic and  melodic interval and temporal metrics, since the original piece was very consonant and had a clear melody.  Unlike the  FHMM, which also had a hierarchical structure, the  LHMM was able to model harmonies  better than the HMM (\autoref{twinkle_layered}).  

After the lowest level in the  LHMM stack, the observed notes were hidden states found by the Viterbi algorithm.  This allowed the second and third layers to explicitly model the harmonies, which led to more consonant pieces and a better modeling of the harmonies in the original piece.  Furthermore, the  LHMM appeared to be able to capture melodic aspects of the hidden states at higher levels, leading to generated pieces with better temporal structure than the HMM.  The hidden states themselves were encoding some aspects of melody, as melody and harmony are very related, and the  LHMM seemed to be able to capture additional aspects of the melody by treating the most-likely hidden states as observed states for higher layers.

\begin{figure}[tbp]
\begin{center}
\includegraphics[width=\textwidth]{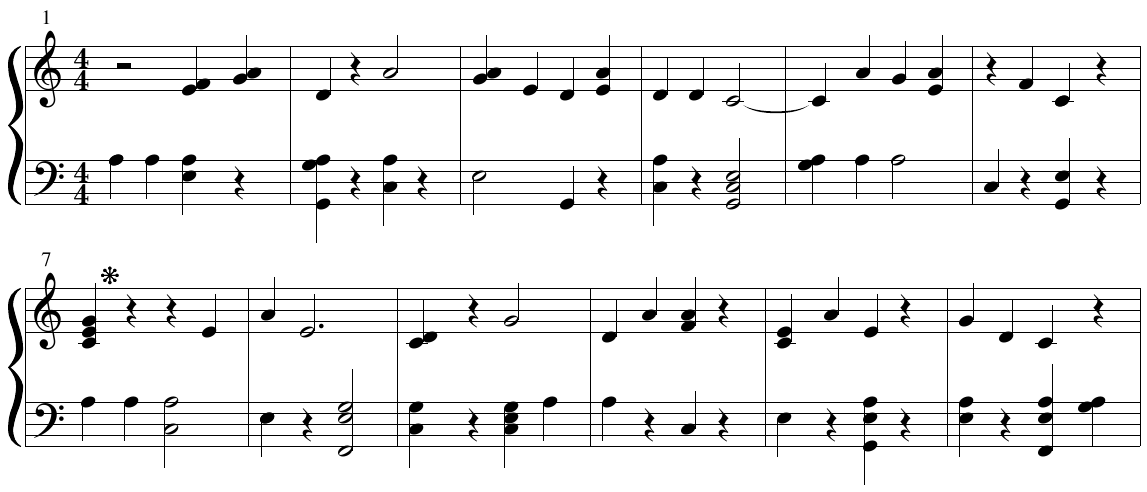}
\end{center}
\caption{Sheet music for a piece generated by a TVAR model with order 7.}
\label{twinkle_TVAR}
\end{figure}

\paragraph{TVAR}
The TVAR model had the highest order dependence between states of any of the models, and these additional dependencies improved the melodic aspects of the generated pieces (\autoref{twinkle_TVAR}).  The TVAR model tended to perform among the best in terms of the temporal metrics.  However, the TVAR model also generated more dissonant pieces than the HMM and did not seem  to capture the harmonic aspects of the original piece very well.  Like with the TSHMMs, the TVAR modeled the hidden process with multiple components (in this case a vector) and the decomposition of the harmonic modeling into multiple components did not appear to improve the modeling of the harmonies in practice. The TVAR model, however, was the only non-stationary model considered, so for more complicated pieces where the key or harmonies are modulated, the TVAR model may be able to outperform the stationary HMM models.

\paragraph{Summary}
Overall, the models considered reveal that additional dependence in the hidden states only does not improve the modeling of the harmonies or the melody in the original piece.  However, overall, melody is not very well-captured in any of the models considered.  One layer of hidden states gives the best modeling of the primary harmonies in a piece and adding additional components to the hidden states hinders the modeling of harmony.  Additional hidden layers or hierarchy did tend to capture more aspects of melody and led to generated pieces with more temporal structure.  For each model, the hidden states modeled the main harmonies in the original piece.  However, for more complex pieces with changing harmonies and key modulations, the stationarity assumptions of the HMM may cause harmony to be less well modeled and lead to more dissonant generated pieces.    

For the example of Twinkle, Twinkle, Little Star, the ideal model seems to be a hierarchical model working with the hidden states explicitly, such as the  LHMM, with high-order lagged dependencies, like the TVAR model.

\section{Listening Evaluation Discussion}

Additional comments provided during the human listening evaluation are included in \autoref{comments}.

\begin{table}[h!]
\scriptsize
\centering
\caption[]{\raggedright Listener comments by the two listening groups on each of the three generated pieces.}
\begin{tabular}{ |p{1.7cm}||p{6.2cm}|p{6.2cm}|  }
 \hline
& \multicolumn{2}{c|}{Comments}\\ \hline
Generated Piece & Musical Ensemble Group & Non-Musical Ensemble Group \\
 \hline\hline
Layered \mbox{HMM -} Chopin's Marche \mbox{funebre}  &  - Melody on top of bass line added complexity  \newline - Movement between octaves made piece sound human composed   \newline - Dissonance in piece more resolved than in other pieces  \newline - ``Somewhat" of a melody that built towards end  \newline - Sounded like ``funeral" piece or Chopin nocturne        &         
 - Half thought complex piece, melodic with a few different themes, more likely to be human-composed  \newline - Half thought ``rote", repetitive, less likely to be human-composed  \newline - Abrupt ending  \newline - Lack of dissonance  \newline - Piece sounded modern or like Chopin\\\hline

Layered \mbox{HMM -} Mendelssohn's Hark! The Herald \mbox{Angels} Sing &    - Too many half-steps / dissonances to sound like human-composed \newline - Too simple, predictable and ``choppy"    \newline - 5/8 thought piece sounded like original training piece, more missed notes and dissonance    &   - Unintended, random dissonance, less likely to be human-composed (or in style of Modern composer)  \newline - Note progressions sounded like human composition  \newline - 1/8 thought piece sounded like original training piece  \\\hline

\mbox{First Order} \mbox{HMM -} Beethoven's Ode to Joy &   - Repetitive melodies, lower voice ``boring"  \newline - Piece could be developed into larger work, used as interlude music  \newline - Atonal, out-of-place dissonance, sounded like a Modern composition  \newline - Better phrases, melodic progression that other pieces, ``good structure" to piece         &           - ``Jarring" dissonance, less so than some of previous pieces  \newline - Enjoyed depth/texture, faster tempo  \newline - Sounded like human-composed ``New-Age" piano music  \newline - Repetitive, piece ``forgot" what had previously occurred and repeated itself  \newline - Sounded like Christmas music, too dissonant to be human-composed\\\hline
\end{tabular}
\label{comments}
\end{table}

\end{document}